\documentclass[authoryear,preprint,review,12pt]{elsarticle_arxiv}

\usepackage{amssymb,amsmath}
\usepackage{bm}
\journal{Icarus}

\begin{document}

\begin{frontmatter}

  \title{Systematic ranging and late warning asteroid impacts}

  \author[jpl]{D. Farnocchia}
  \ead{Davide.Farnocchia@jpl.nasa.gov}
  \author[jpl]{S.~R. Chesley}
  \author[esa,spacedys,iaps]{M. Micheli}
  \address[jpl]{Jet Propulsion Laboratory, California Institute of
    Technology, Pasadena, CA 91109, USA}
  \address[esa]{ESA NEO Coordination Centre, 00044 Frascati (RM), Italy}
  \address[spacedys]{SpaceDyS s.r.l., 56023 Cascina (PI), Italy}
  \address[iaps]{INAF-IAPS, 00133 Roma (RM), Italy}

\begin{abstract}
  We describe systematic ranging, an orbit determination technique
  especially suitable to assess the near-term Earth impact hazard
  posed by newly discovered asteroids. For these late warning cases,
  the time interval covered by the observations is generally short,
  perhaps a few hours or even less, which leads to severe degeneracies
  in the orbit estimation process. The systematic ranging approach
  gets around these degeneracies by performing a raster scan in the
  poorly-constrained space of topocentric range and range rate, while
  the plane of sky position and motion are directly tied to the
  recorded observations. This scan allows us to identify regions
  corresponding to collision solutions, as well as potential impact
  times and locations.  From the probability distribution of the
  observation errors, we obtain a probability distribution in the
  orbital space and then estimate the probability of an Earth
  impact. We show how this technique is effective for a number of
  examples, including 2008~TC$_3$ and 2014~AA, the only two asteroids
  to date discovered prior to impact.
\end{abstract}

\begin{keyword}
  Asteroids \sep Asteroids, dynamics \sep Astrometry \sep Near-Earth
  objects \sep Orbit determination
\end{keyword}

\end{frontmatter}

\section{Introduction}
When an asteroid is first discovered, the few available observations
rarely allow the determination of its orbit. Surveys usually capture
3--5 repeat images of the same area of sky with a typical return time
of 15--30 minutes. Over this time interval, an object in the field of
view moves with respect to background stars at a roughly linear
rate. Thus, the short arc of available data, also called a tracklet
\citep{kubica}, provides an estimate of the angular position and
motion of the object in the plane of sky, which is not enough to
determine a six-parameter orbit \citep{milani05}.

Despite the degeneracies in the orbit estimation process, it is
important to recognize potentially hazardous objects shortly after
discovery. In fact, the only two objects discovered prior to an Earth
impact, namely 2008~TC$_3$ and 2014~AA, were both discovered by
R. Kowalski \citep{mpec_tc3, mpec_aa} of the Catalina Sky Survey
\citep{catalina} only about 20 hours before striking the Earth. On one
hand, 2008~TC$_3$ was quickly recognized as potential impactor and so
it was extensively observed before the impact. The acquired  observations
allowed the estimation of the orbit of 2008~TC$_3$ as well as some
physical characterization \citep{jenniskens_nat, kozubal,
  scheirich}. Using the available data, it was possible to predict the
impact that took place above Sudan on 2008 October
7\footnote{http://neo.jpl.nasa.gov/news/2008tc3.html}
\citep{jenniskens_nat}. On the other hand, 2014~AA was not immediately
recognized as a possible impactor and so only seven astrometric
observations over about 70 minutes were obtained before the object
fell into the Atlantic Ocean on 2014 January
2\footnote{http://www.nasa.gov/jpl/asteroid/first-2014-asteroid-20140102}
\citep{chesley_ieee}.

When the standard differential correction procedure
\citep[e.g.,][]{farnocchia_ast4, orbdet} to find a least-squares orbit
fails, other methods can be used to assess the orbital probability
distribution. In particular, \citet{bayes} show how to put a
probability density on the phase space of the orbital elements by
using Bayesian inversion theory. By using a Monte Carlo approach one
can sample the orbit phase space and thereby derive the probability
distribution from the observation errors corresponding to the sampled
orbit.

The available observations directly constrain the position and motion
of the asteroid in the sky while the distance between the asteroid and
the observer (topocentric range) is poorly constrained. Thus, ranging
methods are the preferred Monte Carlo approach when only a short arc
of observations is available. \citet{stat_ran} describe a method
called statistical ranging that allows one to generate orbital samples
compatible with the observational data. After choosing two
observations, the topocentric ranges at the epochs of the two
observations are randomly sampled and a corresponding orbit
computed. \citet{mcmc} rely on the statistical ranging technique and
use Markov chains to generate an unbiased sequence of orbital samples
distributed according to the probability distribution coming from
Bayesian inversion theory.

\citet{sys_ran_steve} introduces a technique called systematic
ranging\footnote{This technique was actually introduced by Tholen and
  Whiteley in 2002, but the paper was never published.}, which in
contrast to Monte Carlo techniques systematically explores a raster in
the topocentric range and range-rate space. This technique provides a
geometric description of the orbital elements as a function of range
and range rate. Moreover, systematic ranging allows one to
identify regions of the phase space filled with impact solutions and
the corresponding impact times and locations. In this paper we present
a detailed description of systematic ranging and show how to
derive a probability distribution on the range and range-rate space,
which is then mapped to the orbital element space where 
impact probability estimates can be derived.

\section{Systematic ranging}
Systematic ranging relies on the fact that a short arc of observations
yields a direct estimate of the plane of sky position (right ascension
$\alpha$ and declination $\delta$) and motion ($\dot\alpha$ and
$\dot\delta$). These four scalar quantities can be assembled together
in the so-called attributable
${\cal A} = (\alpha, \delta, \dot\alpha, \dot\delta)$
\citep{milani05}. The topocentric range $\rho$ and $\dot\rho$ are only
marginally constrained, if at all. If $\rho$ and $\dot\rho$ were
known, we would have a full description of the asteroid's topocentric
position and velocity in polar coordinates
$(\alpha, \delta, \dot\alpha, \dot\delta, \rho, \dot\rho)$, which can
be easily converted to a Cartesian heliocentric state if the position
and velocity of the observer are known. Note that, to account for
light-time correction, we correct the epoch of the Cartesian state
from that of the attributable by a quantity $\rho/c$, where $c$ is the
speed of light.

To explore the orbit phase space, systematic ranging scans a
suitably dense grid in the $(\rho,\dot\rho)$ space. Such grid needs to
be large enough to contain the reasonably possible orbital configurations. In
particular, the grid must contain the so-called admissible region
\citep{milani04}, i.e., the values of $(\rho, \dot\rho)$ leading to
bounded heliocentric orbits. For each grid point we fix the values of
$\rho = \rho_i$ and $\dot\rho = \dot\rho_j$ and find the best fit
value of the attributable ${\cal A}_{ij}$ that minimizes the cost function:
\begin{equation}\label{e:target}
Q = \bm\nu^T W \bm \nu
\end{equation}
where $\bm\nu$ is the vector of Observed $-$ Computed astrometric
residuals and $W$ is the weight matrix
\citep{farnocchia_ast4}. 

The constrained best-fitting solution can easily be converted to an
orbit, which is in turn propagated to find upcoming Earth
encounters. Moreover, if the observations contain photometric
measurements we also compute the absolute magnitude for
each grid point.

\begin{figure}
\centerline{\includegraphics[width=0.5\textwidth]{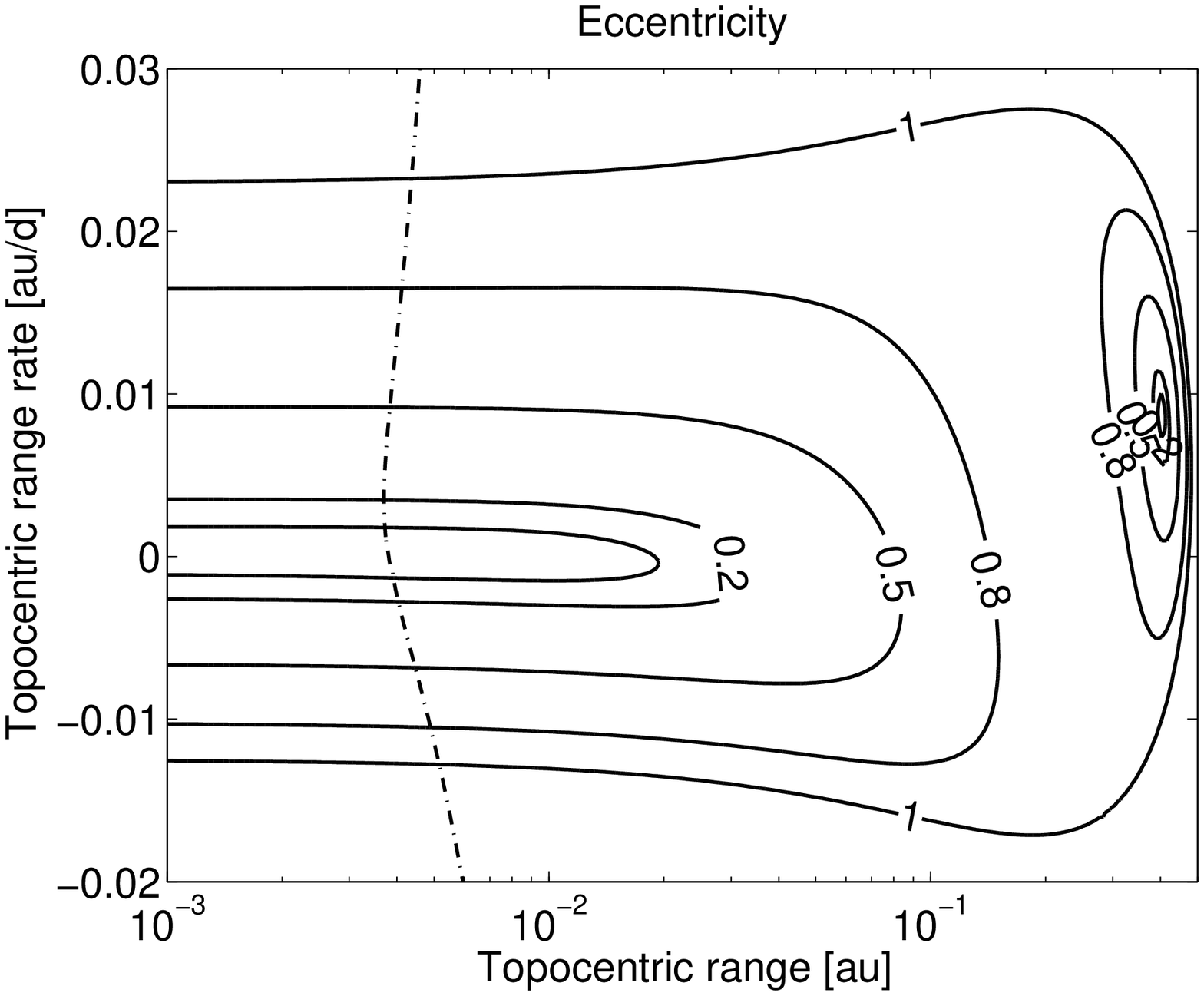}
\includegraphics[width=0.5\textwidth]{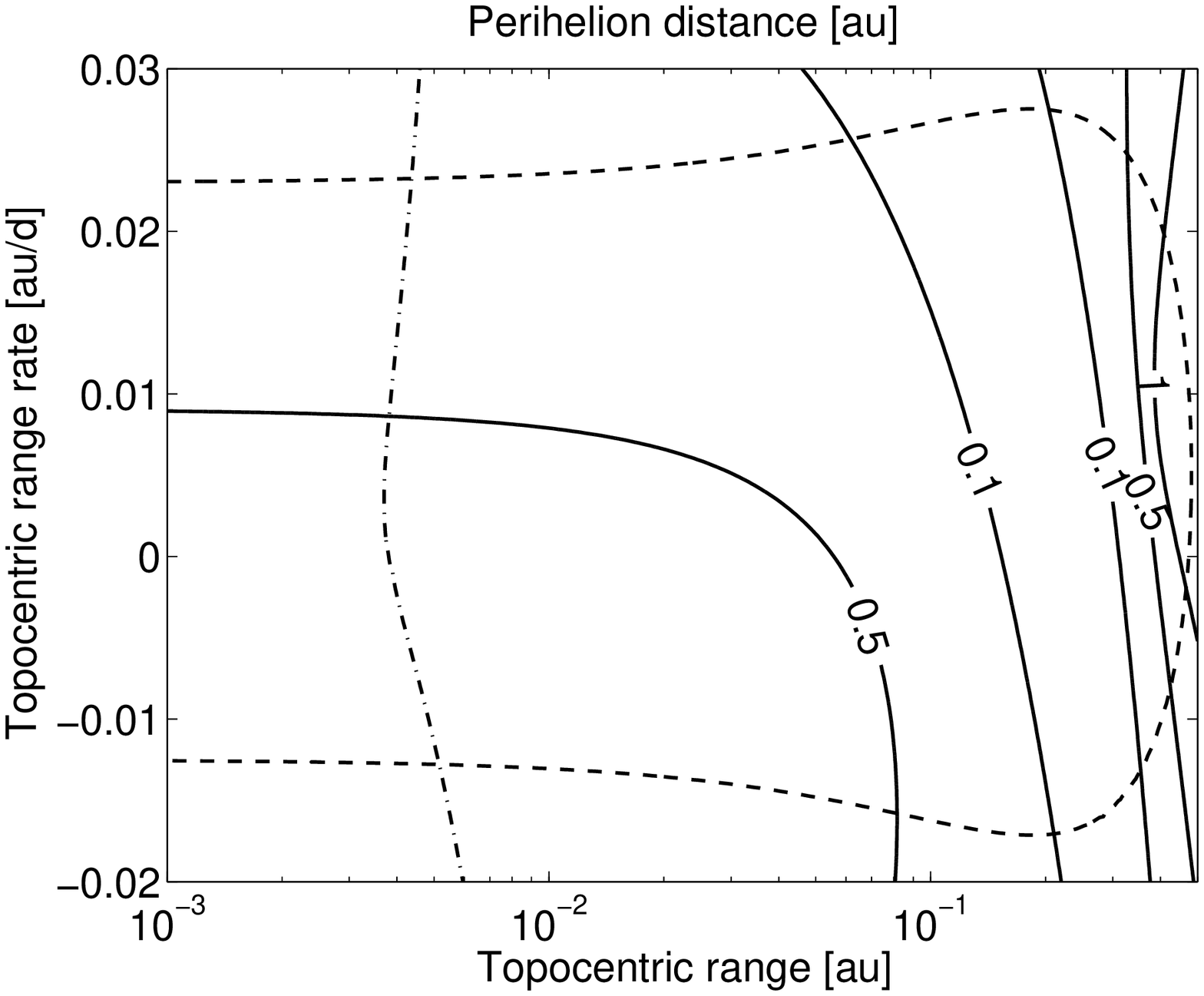}}
\centerline{\includegraphics[width=0.5\textwidth]{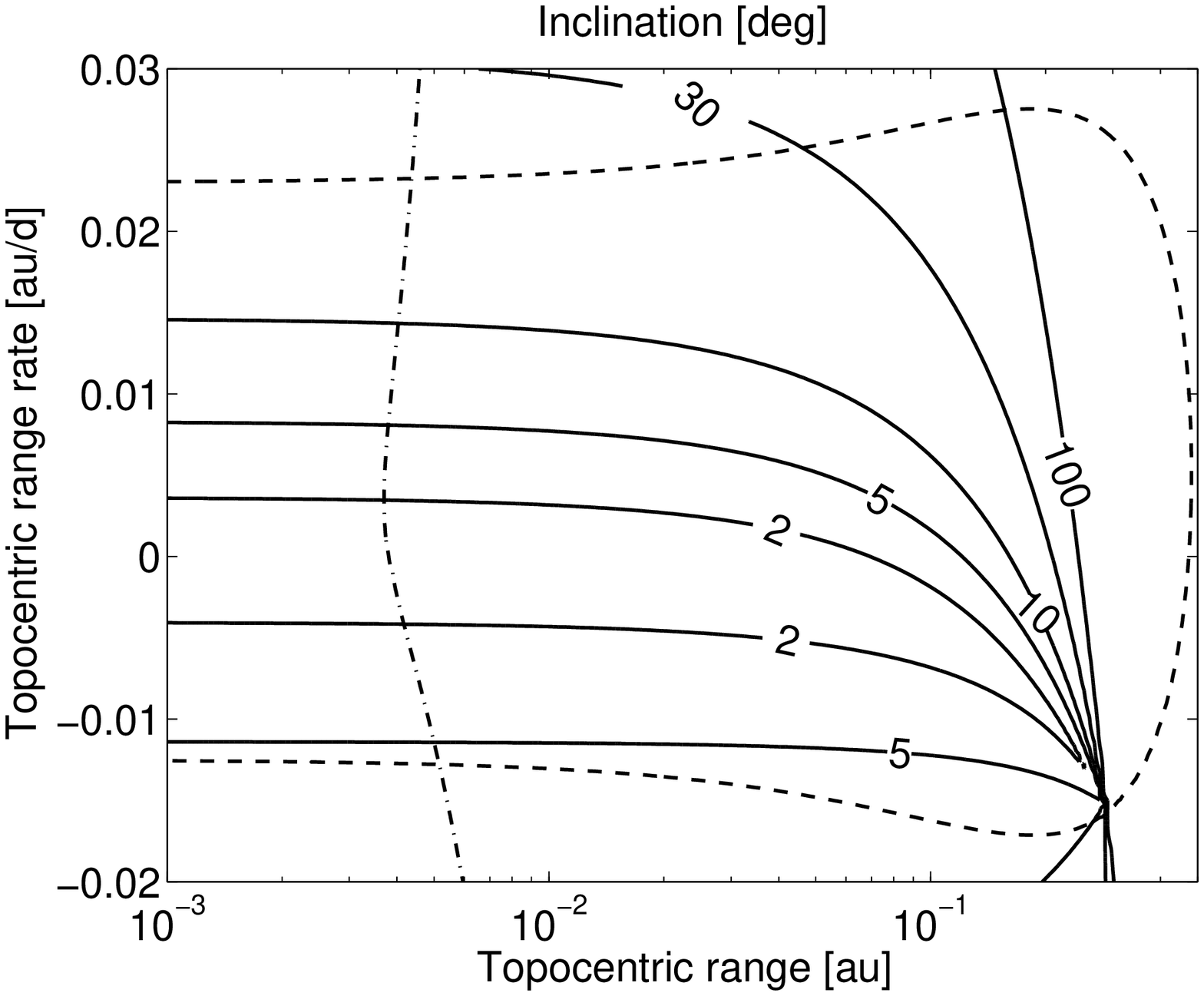}
\includegraphics[width=0.5\textwidth]{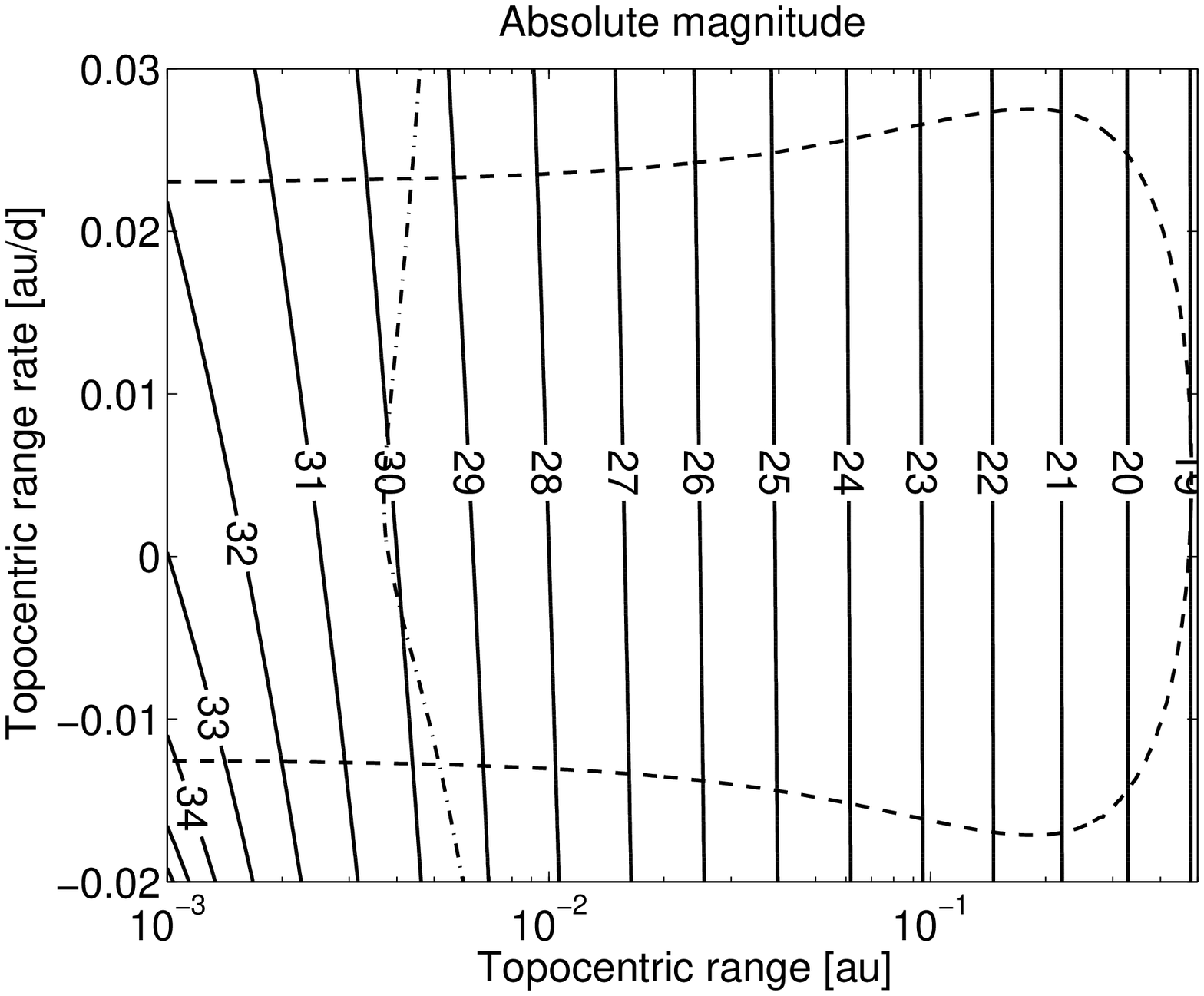}}
\caption{Eccentricity, perihelion distance, inclination, and absolute
  magnitude as a function of range and range rate for asteroid
  2014~AA. The dashed curve corresponds to parabolic orbits and the
  dash-dotted line to grazing impacts.}\label{f:2014aa}
\end{figure}

Figure~\ref{f:2014aa} shows the application of systematic ranging
to 2014~AA by using the seven astrometric observations obtained by the
Catalina Sky Survey prior to impact. The orbital elements are shown as
a function of $(\rho, \dot\rho)$. The dashed curve encloses bounded
orbits and corresponds to the admissible region of \citet{milani04}.
The dash-dotted line is for grazing impacts: the region on the left of
the curve contains impacting solutions. 

\section{Probability distribution on the range \& range rate
  space}\label{s:rrd_dist}
The raster scan presented in the previous section gives a geometric
description of how the orbital configuration depends on the value of
topocentric range and range rate. The next step is to assign a
probability distribution to this space.

As is common practice, given a set of optical observations we assume
that the observation errors $\bm\nu$ are normally distributed
according to a weight matrix $W$, i.e., their probability density
$f_{err}$ is:
\[
f_{err}(\bm\nu) \propto \exp(-0.5\ \bm\nu^T\,W\,\bm\nu)\ .
\]
It is typical to use a diagonal weight matrix where the individual
weights are $1/\sigma^2$ ($\sigma$ is the uncertainty in the
astrometric positions). Table~\ref{t:weights} shows the data weights
we currently use for the most productive discovery and follow-up
stations. The chosen weights are based on our experience and account
for the fact that discovery observations can sometimes be problematic
because they are not targeted on the object.


\begin{table}[t]
\begin{center}
\begin{tabular}{|cc|cc|cc|}
  \hline
  Station & $\sigma_{\alpha,\delta}$ & Station &
                                                 $\sigma_{\alpha,\delta}$
  & Station & $\sigma_{\alpha,\delta}$\\
  \hline 
  703 & 1.00$''$ & G96 & 0.50$''$ & F51 & 0.20$''$ \\
  950 & 0.50$''$ & 291 & 0.70$''$ & 691 & 0.70$''$ \\
  568 & 0.15$''$ & H01 & 0.30$''$ & H21 & 0.70$''$ \\
  J04 & 0.40$''$ & W84 &  0.20$''$ & F65, E10 & 0.40$''$\\
  W85, W86, W87 & 0.60$''$ & Q63, Q64, V37 & 0.80$''$ &  K91, K92, K93
  & 0.80$''$\\
\hline
\end{tabular}
\end{center}
\caption{Station-specific weighting rules for the most productive
  discovery and short-term follow-up stations. For all the other
  stations we use weights at 1$''$.}
\label{t:weights}
\end{table}

In principle, it is possible to consider the photometric residuals
together with the astrometric ones. However, photometric measurements
have a much larger uncertainty than that of the astrometry and can be
affected by significant biases \citep{sdss}. Moreover, in case of a
magnitude trend, it is hard to tell if the varying luminosity is due
to a rapidly changing topocentric distance rather than the asteroid's
rotation. Therefore, we do not use the information obtained from the
photometric residuals.

According to Bayesian inversion theory \citep{bayes}, the
posterior probability density function for
$(\rho,\dot\rho)$ is:
\[
f_{post}(\rho,\dot\rho) \propto f_{err}(\bm\nu(\rho, \dot\rho))\,
f_{prior}(\rho,\dot\rho)\ ,
\]
where $f_{prior}$ is a prior distribution on the $(\rho, \dot\rho)$
space.\footnote{It is often convenient to use a logarithmic scale for
  $\rho$ to achieve a better resolution at small topocentric
  distances. In that case the probability density has to be multiplied
  by $\rho$.} The selection of $f_{prior}$ is to some extent arbitrary
and far from trivial \citep{prior}, and yet can significantly affect
the posterior probability distribution, especially for short
observation arcs.

Whatever the choice of $f_{prior}$, we always add a crude constraint
from the population model by setting $f_{prior} = 0$ for hyperbolic
orbits, flagging objects where the astrometric errors favor an unbounded
orbit.

\subsection{Jeffreys' prior}
A mathematically sound choice is Jeffreys' prior \citep{openorb}:
\begin{equation}\label{e:jeffreys}
f_{prior}(\rho, \dot\rho) = \sqrt{\det\left(\frac{\partial
  \bm\nu}{\partial (\rho, \dot\rho)}^T\  W \ \frac{\partial
  \bm\nu}{\partial (\rho, \dot\rho)}\right)}\ .
\end{equation}
Note that the partials are the total derivatives of $\bm \nu$ with
respect to $(\rho, \dot\rho)$, which means that they account for
the fact that the best-fit attributable ${\cal A}$ changes as a
function of $(\rho, \dot\rho)$.

Among other things, Jeffreys' prior secures the invariance of the
probability distribution when changing variables. However, since there
is more sensitivity of the residuals for small topocentric distances,
Jeffreys' prior tends to favor orbital configurations where the object
is close to the observer, e.g., see the examples of 2015~CV and
2015~BU$_{92}$ in Sec.~\ref{s:examples}.

\subsection{Prior based on population model}
As already discussed by \citet{stat_ran}, the prior could be based on
the known asteroid population. Even better, one could use an orbit and
size population model where the known population is extrapolated to
remove survey selection effects and to reach completeness up to a
given apparent magnitude.  One such model is the Pan-STARRS Synthetic
Solar System Model \citep{grav11}, and new population models are
expected in the future \citep{granvik_neo}. From the \citet{grav11}
model we derived a probability density $f_{pop}(q, e, i, H)$ for
perihelion distance $q$, eccentricity $e$, inclination $i$, and
absolute magnitude $H$. A more refined approach would be to remove the
known population from the population model.

Mapping $f_{pop}$ to the best fitting orbits in the
$(\rho,\dot\rho)$ plane is quite complicated. Therefore, to compute such a
prior we iteratively apply Bayesian inversion theory, i.e., the
prior is computed as posterior $f_{post}'$ of another prior
$f_{prior}'$:
\[
f_{prior}(\rho, \dot\rho) = f_{post}' (\rho, \dot\rho) \propto f_{prior}'
(\rho, \dot\rho) f_{pop}(q (\rho, \dot\rho), e (\rho, \dot\rho), i
(\rho, \dot\rho), H (\rho, \dot\rho))\,.
\]
At this point we select $f_{prior}'$ using simple geometric
considerations. By assuming that the Cartesian space of position and
velocity is uniformly filled, we have that $\rho^3$ and $\dot\rho$ are
uniformly distributed. Hence,
$f_{prior}' (\rho, \dot\rho) \propto \rho^2$.

\subsection{Uniform prior}
A common choice for $f_{prior}$ is a uniform distribution. For
instance, such a prior is used by \citet{stat_ran} and \citet{mcmc}.
Interestingly, it is possible to approximately obtain a prior
distribution from the population model if we only consider the size
distribution $f_{size}(H)$, i.e., $f_{prior} = \rho^2 f_{size}(H) $.

For near-Earth asteroids the cumulative size
distribution is expected to follow a power law
$f_{size}(H) \propto 10^{\eta H}$, with $\eta$ somewhere between 0.35
and 0.47 \citep{bottke02, harris, stuart}. For a given visual
magnitude $V$, we have that $f_{size}(H) \propto \rho^{-5\eta}$
\citep{bowell} and therefore $f_{prior} = \rho^{2 - 5\eta}$. According
to the range of possible values for $\eta$, this prior corresponds to
a power law with exponent between $-0.35$ and $0.25$, which is
compatible with a uniform distribution.

Though a uniform distribution might appear overly simplistic, we have found
that this it does a good job of identifying potential
impactors (e.g., see Sec.~\ref{s:examples}) and distinguishing
interesting objects from more common ones, such as main belt asteroids.

\section{Impact probability computation}\label{s:imp_prob}
For each grid point, as a result of the constrained least squares fit
to the observations, we obtain not only a best fit value
${\cal A}_{ij}$ of the attributable, but also the corresponding
uncertainty expressed by a covariance matrix $\Gamma_{{\cal
    A}_{ij}}$.
This uncertainty is easily mapped to the orbital element space and
therefore allows us to compute the probability $p_{ij}$ of an Earth
impact for that orbit. Since for a fixed grid point the orbital
uncertainty is small and the horizon for impacts is of the order of
weeks, we compute the impact probability with a linear approach
\citep{milani_ast3, farnocchia_ast4}.

Given $f_{post}$ we can assign each grid point a weight
$w_{ij} \propto f_{post}(\rho_i, \dot\rho_j)$, where
$\sum w_{ij} = 1$. Then, the impact probability can be easily computed
as
\[
P = \sum w_{ij} p_{ij}.
\]

\section{Generation of Monte Carlo orbital samples}
To better visualize the orbital distribution, it is convenient to
generate Monte Carlo orbital samples distributed according to the
probability distribution computed above. The following procedure can
be used to achieve this goal:
\begin{description}
\item[(a)] Randomly draw $\rho_k$ according to a uniform distribution
  between the minimum and maximum values of $\rho$ on the grid;
\item[(b)] Randomly draw $\dot\rho_k$ according to a uniform distribution
  between the minimum and maximum values of $\dot\rho$ on the grid;
\item[(c)] Randomly draw a value $\gamma$ according to a uniform
  distribution between 0 and the maximum of $f_{post}$ on
  the grid;
\item[(d)] Compute a best-fit attributable ${\cal A}_k$ for
  $(\rho, \dot\rho) = (\rho_k, \dot\rho_k)$;
\item[(e)] If $f_{post}(\rho_k, \dot\rho_k) \geq \gamma$ go
  to (f), otherwise start again from (a);
\item[(f)] Add normally distributed noise to ${\cal A}_k$ according to
  the covariance $\Gamma_{{\cal A}_k}$ and add
  $({\cal A}_k, \rho_k, \dot\rho_k)$ to the sequence of orbital
  samples.
\end{description}

\section{Examples}\label{s:examples}
We want to test the algorithm presented in the previous sections on
real cases. On one hand it is important that systematic ranging
provides an early detection of impacting asteroids. On the other hand
we need to avoid generating too many warnings for objects that might
be very far from Earth. Thus, we show examples of impactors as well as
main belt asteroids. Table~\ref{t:examples} summarizes the results for
the examples discussed in more detail in the following subsections.

\begin{table}[t]
\small
\begin{center}
\begin{tabular}{|lccc|ccc|}
  \hline
  Object & Number & Arc & Weights & \multicolumn{3}{c|}{Impact probability}\\
 & of obs & length & & Jeffreys & Uniform & Population\\
  \hline 
  2008~TC$_3$ & 4 & 43 min & 0.5$''$ & $6.3\times 10^{-1}$ & $4.4\times 10^{-2}$
 & $2.4\times 10^{-2}$\\
  2008~TC$_3$ & 4 & 43 min & 1.0$''$ & $6.7\times 10^{-1}$ & $3.7\times 10^{-3}$
 & $3.5\times 10^{-3}$\\
  2008~TC$_3$ & 7 & 99 min & 0.5$''$ & $1.0$ & $1.0$
 & $1.0$\\
  2008~TC$_3$ & 7 & 99 min & 1.0$''$ & $9.7\times 10^{-1}$ & $9.8\times 10^{-1}$
 & $9.6\times 10^{-1}$\\
  \hline
  2014~AA & 3 & 28 min & 0.5$''$ & $8.1\times 10^{-1}$ & $2.9\times 10^{-2}$
 & $3.1\times 10^{-2}$\\
  2014~AA & 3 & 28 min & 1.0$''$ & $8.4\times 10^{-1}$ & $3.3\times 10^{-3}$
 & $8.4\times 10^{-3}$\\
  2014~AA & 7 & 69 min & 0.5$''$ & $1.0$ & $1.0$
 & $1.0$\\
  2014~AA & 7 & 69 min & 1.0$''$ & $9.9\times 10^{-1}$ & $9.6\times 10^{-1}$
 & $9.0\times 10^{-1}$\\
  \hline
  Hayabusa & 3 & 7 min & 1.0$''$ & $9.4\times 10^{-1}$ & $2.7\times 10^{-3}$
 & $2.0\times 10^{-3}$\\
  Hayabusa & 6 & 34 min & 1.0$''$ & $2.4\times 10^{-1}$ & $1.7\times 10^{-1}$
 & $2.2\times 10^{-2}$\\
  \hline
  Hayabusa 2 & 3 & 43 min & 0.2$''$ & $2.8\times 10^{-1}$ & $3.6\times 10^{-1}$
 & $4.2\times 10^{-1}$\\
  Hayabusa 2 & 3 & 43 min & 0.5$''$ & $3.6\times 10^{-1}$ & $3.3\times 10^{-2}$
 & $7.2\times 10^{-2}$\\
  Hayabusa 2 & 6 & 26 h & 0.2$''$ & $2.4\times 10^{-13}$ & $2.3\times 10^{-13}$
 & $2.2\times 10^{-13}$\\
  Hayabusa 2 & 6 & 26 h & 0.5$''$ & $1.9\times 10^{-3}$ & $1.9\times 10^{-3}$
 & $1.8\times 10^{-3}$\\
  \hline
  2004~AS$_1$ & 4 & 71 min & 1.0$''$ & $2.2\times 10^{-1}$ & $1.2\times 10^{-1}$
 & $7.8\times 10^{-2}$\\
  2004~AS$_1$ & 4 & 71 min & 1.5$''$ & $1.2\times 10^{-1}$ & $1.3\times 10^{-2}$
 & $1.1\times 10^{-2}$\\
  \hline
  2004~FU$_{162}$ & 4 & 44 min & 1.0$''$ & $3.6\times 10^{-1}$ & $7.3\times 10^{-3}$
 & $8.7\times 10^{-4}$\\
  2004~FU$_{162}$ & 4 & 44 min & 1.5$''$ & $5.0\times 10^{-1}$ & $1.3\times 10^{-2}$
 & $1.0\times 10^{-3}$\\
  \hline
  2015~CV & 4 & 34 min & 0.5$''$ & $8.5\times 10^{-3}$ & $4.7\times 10^{-6}$
 & $9.9\times 10^{-7}$\\
  2015~CV & 4 & 34 min & 1.0$''$ & $1.7\times 10^{-1}$ & $1.5\times 10^{-4}$
 & $4.6\times 10^{-5}$\\
  \hline
  2015~BU$_{92}$ & 3 & 36 min & 0.5$''$ & $1.7\times 10^{-1}$ & $7.2\times 10^{-5}$
 & $3.5\times 10^{-10}$\\
  2015~BU$_{92}$ & 3 & 36 min & 1.0$''$ & $4.8\times 10^{-1}$ & $4.0\times 10^{-4}$
 & $8.5\times 10^{-8}$\\
  \hline
\end{tabular}
\end{center}
\caption{Impact probabilities for the examples considered in
  Sec.~\ref{s:examples} for different choices of $f_{prior}$,
  different number of tracklets, and different data weights.}
\label{t:examples}
\end{table}

\subsection{2008~TC$_3$}
\begin{figure}
\centerline{\includegraphics[width=0.5\textwidth]{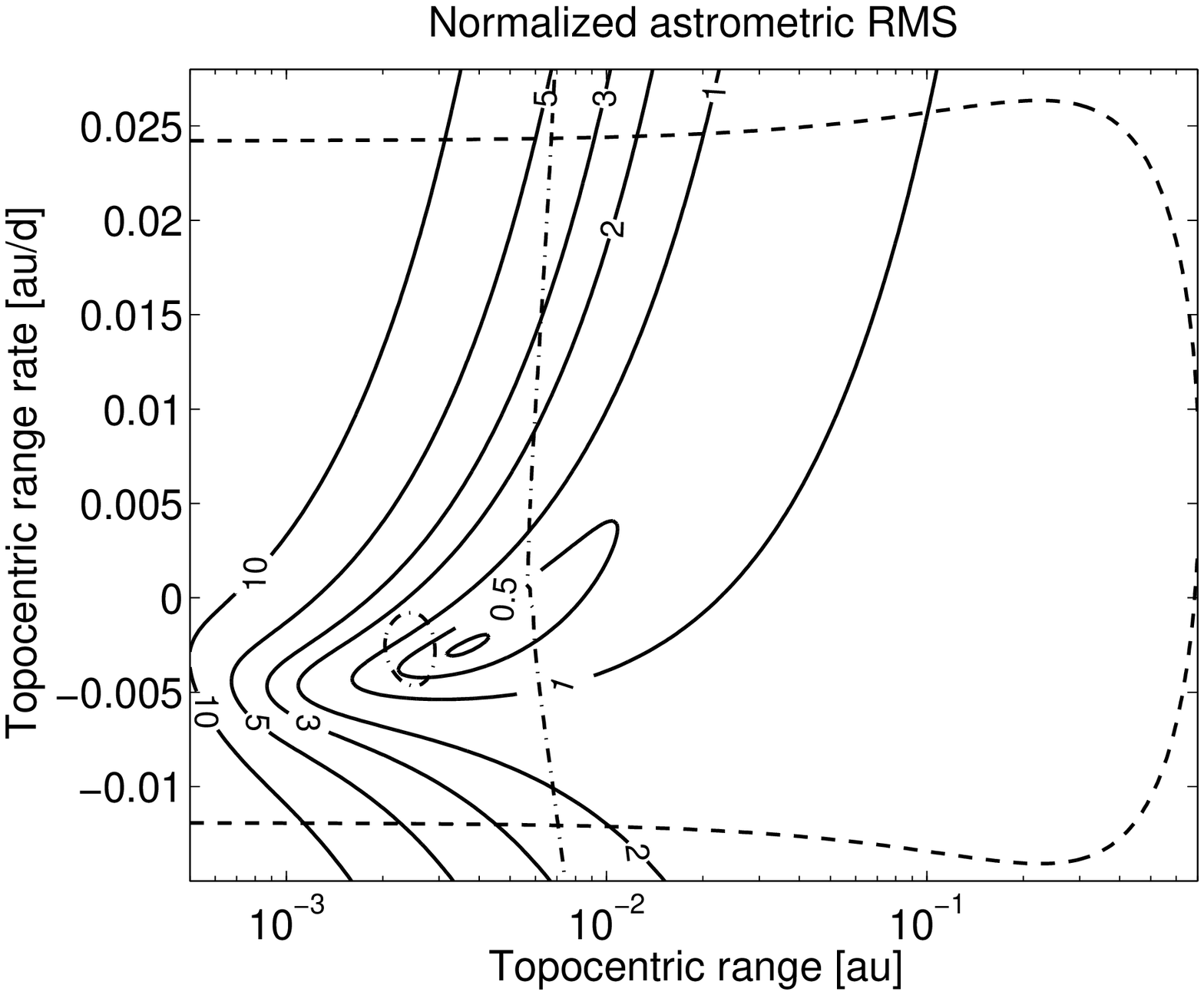}
\includegraphics[width=0.5\textwidth]{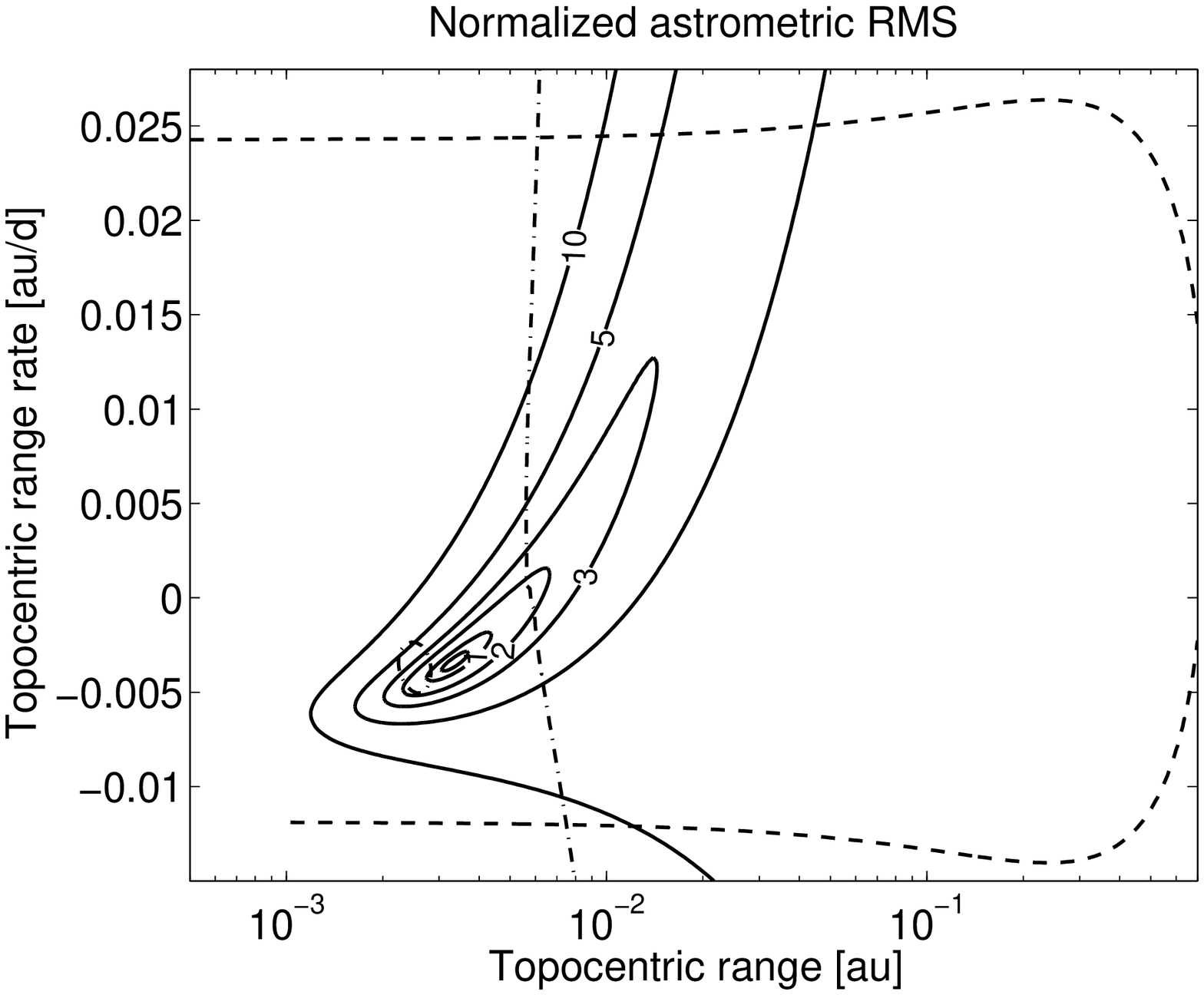}}
\caption{Normalized RMS as a function of $(\rho, \dot\rho)$ for
  2008~TC$_3$. Left panel is for the first tracklet (four observations
  over 43 min), right panel is for the first two tracklets (total of
  seven observations over 99 min). Astrometric weights are at
  0.5$''$. The dashed curve corresponds to parabolic heliocentric
  orbits and the dash-dotted line to grazing
  impacts.}\label{f:tc3_rms}
\end{figure} 
As already discussed in the introduction, 2008~TC$_3$ was the first
asteroid discovered prior to impact. This object was first spotted and
promptly followed up by R.~A. Kowalski at the Mt. Lemmon
observatory. In the 20 hours between discovery and impact hundreds of
astrometric observations were reported to the Minor Planet
Center. These observations allowed the calculation of a least squares
orbit and the prediction of the impact. How would systematic
ranging have performed with the first observations of 2008~TC$_3$?

The left panel of Fig.~\ref{f:tc3_rms} shows the normalized RMS, i.e.,
the square root of the target function $Q$ in Eq.~\eqref{e:target}, as
a function of $(\rho, \dot\rho)$ by using the first tracklet of four
observations obtained by Kowalski. The nonlinearity of the orbit
determination problem is evident from the level curves of the RMS,
which clearly suggest that 2008~TC$_3$ could be on an impact
trajectory. As a matter of fact, the corresponding impact probability,
computed as described in Sec.~\ref{s:imp_prob}, is already significant
with the first four observations (Table~\ref{t:examples}). With our
defaults data weights at 0.5$''$, the impact probability is few
percents when using a uniform prior or one based on the population
model, and $\sim$60\% when using Jeffreys' prior. With more conservative
weights at 1$''$ Jeffreys' prior leads to a $\sim$70\% impact
probability, while the uniform and population prior result in a
probability of few in a thousand.



With the addition of Kowalski's second tracklet, three additional
observations, the orbit becomes much better constrained by the
observation residuals (see right panel of Fig.~\ref{f:tc3_rms}). The
impact is now almost certain, especially with 0.5$''$ weights.

\begin{figure}
\centerline{\includegraphics[width=10 cm]{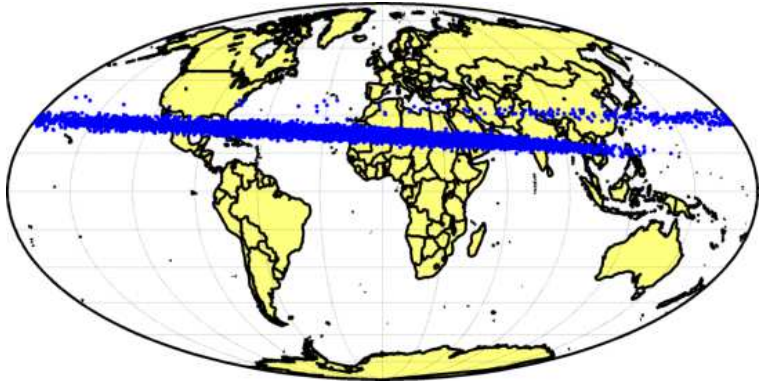}}
\caption{Potential impact locations for 2008 TC$_3$ based on the first 
  two tracklets of data.}\label{f:tc3_loc}
\end{figure}

By using the first two tracklets of data and adopting a uniform prior, we
also generated Monte Carlo samples to find potential impact locations,
as shown in Fig.~\ref{f:tc3_loc}. As expectable, the region of
possible impact location contains the Sudan desert where 2008~TC$_3$
actually ended its journey \citep{jenniskens_nat}.

\subsection{2014~AA}
Though 2014~AA was not extensively followed up after discovery, there
are several similarities with 2008~TC$_3$. In particular, 2014~AA
was also discovered about 20 hours before impact by R.~A. Kowalski at the
Mt. Lemmon observatory.

\begin{figure}
\centerline{\includegraphics[width=0.5\textwidth]{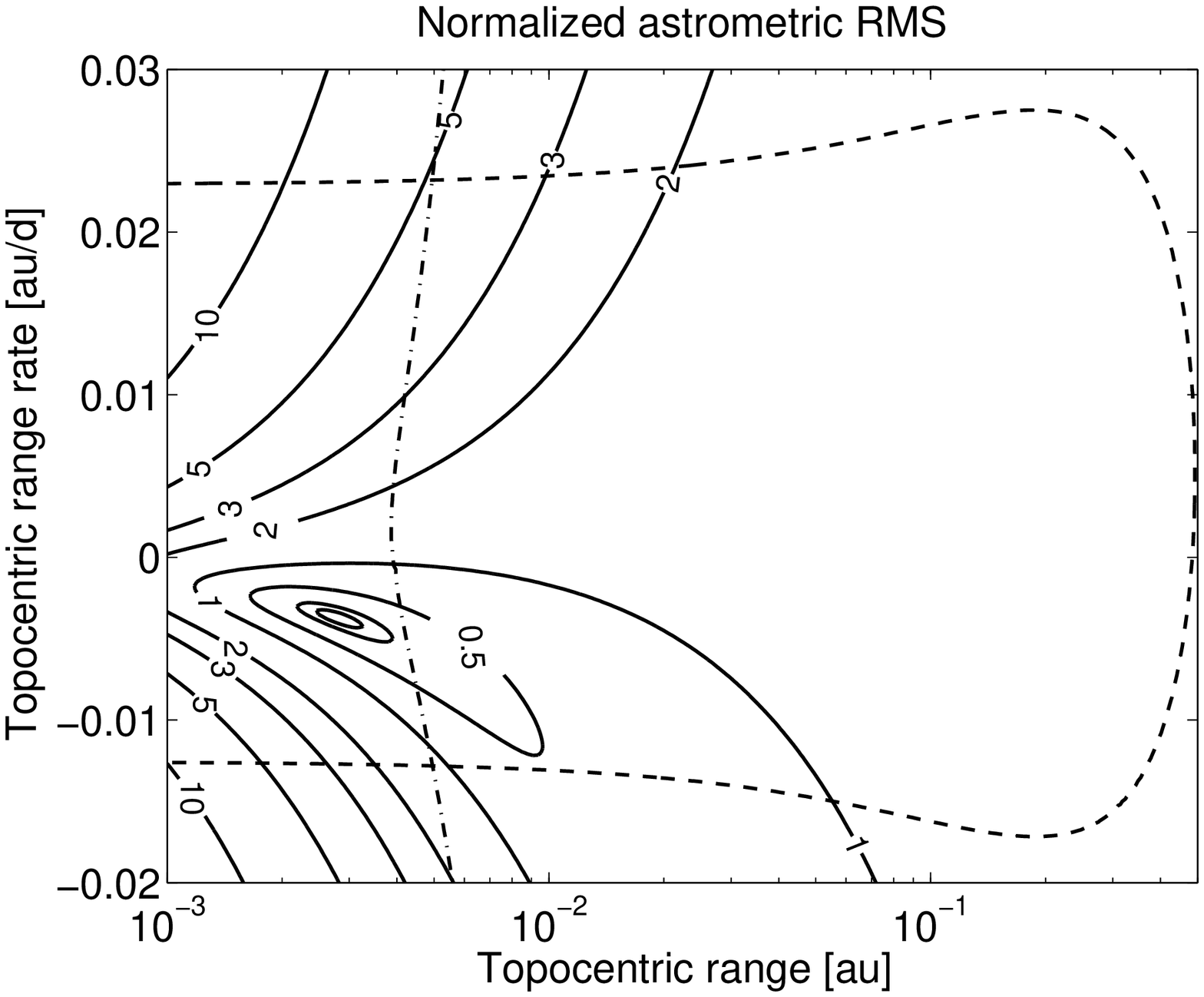}
\includegraphics[width=0.5\textwidth]{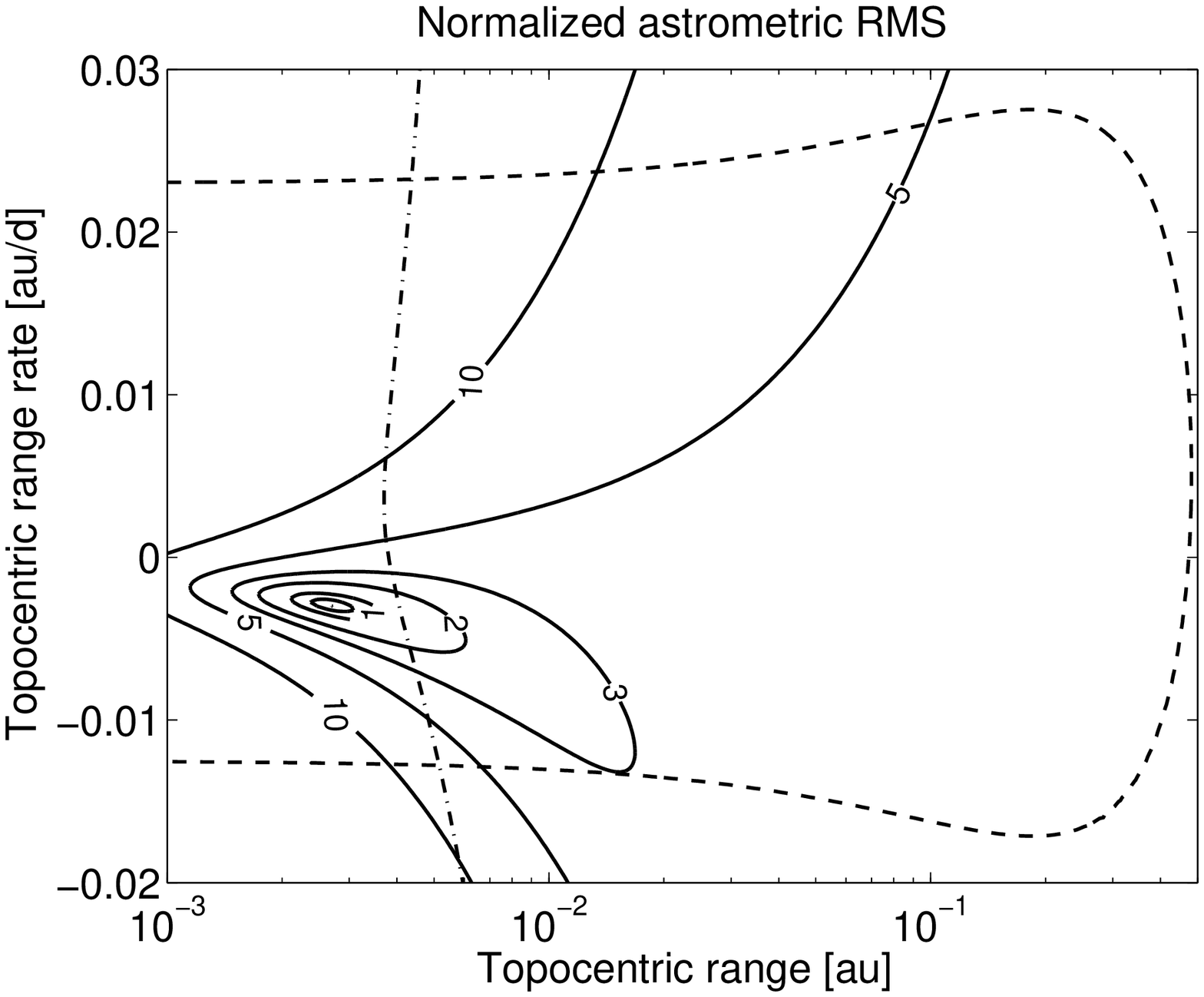}}
\caption{Normalized RMS as a function of $(\rho, \dot\rho)$ for
  2014~AA. Left panel is for the first tracklet (three observations
  over 28 min), right panel is for the first two tracklets (total of
  seven observations over 69 min). Astrometric weights are at
  0.5$''$. The dashed curve corresponds to parabolic heliocentric
  orbits and the dash-dotted line to grazing impacts.}\label{f:aa_rms}
\end{figure}

Figure~\ref{f:aa_rms} shows the normalized RMS for 2014~AA as a
function of $(\rho, \dot\rho)$ by using the first tracklet of data
(three observations) or the full available dataset (seven
observations). Again, the RMS curves clearly indicate that an impact
is possible. Systematic ranging immediately recognizes the potential
impact from the first tracklet and make the impact almost certain when
the second tracklet is added. Table~\ref{t:examples} reports the
computed impact probabilities, which are similar to those computed for
2008~TC$_3$,.



\begin{figure}
\centerline{\includegraphics[width=10 cm]{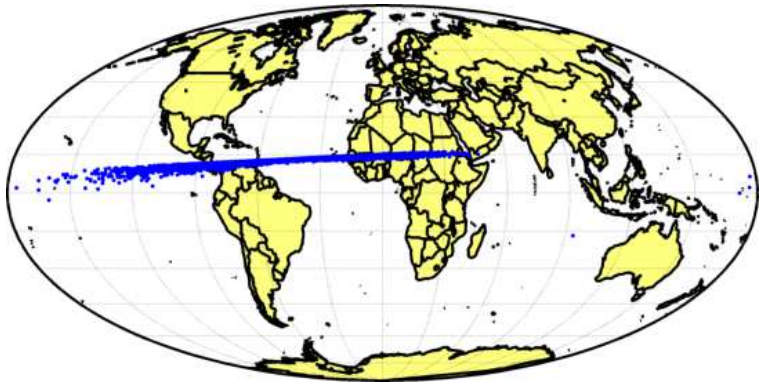}}
\caption{Potential impact locations for 2014~AA based on all available
  data, i.e., seven observations.}\label{f:aa_loc}
\end{figure}
The impact locations (see Fig.~\ref{f:aa_loc}) found by using the
seven available observations and adopting a uniform prior are
statistically consistent with the detection by the infrasound
component of the International Monitoring System (IMS) operated by the
Comprehensive Nuclear Test Ban Treaty Organization (CTBTO), which
places the impact location in the Atlantic Ocean at a latitude of
about 15$^\circ$ N and a longitude of about 43$^\circ$ W
\citep{chesley_ieee,brown_acm}.

\subsection{Hayabusa \& Hayabusa 2}
Because of the low number of actual impactors we can use as a
validation, we tested our algorithm on Hayabusa and Hayabusa 2
\citep{hayabuse}.

\begin{figure}
\centerline{\includegraphics[width=0.5\textwidth]{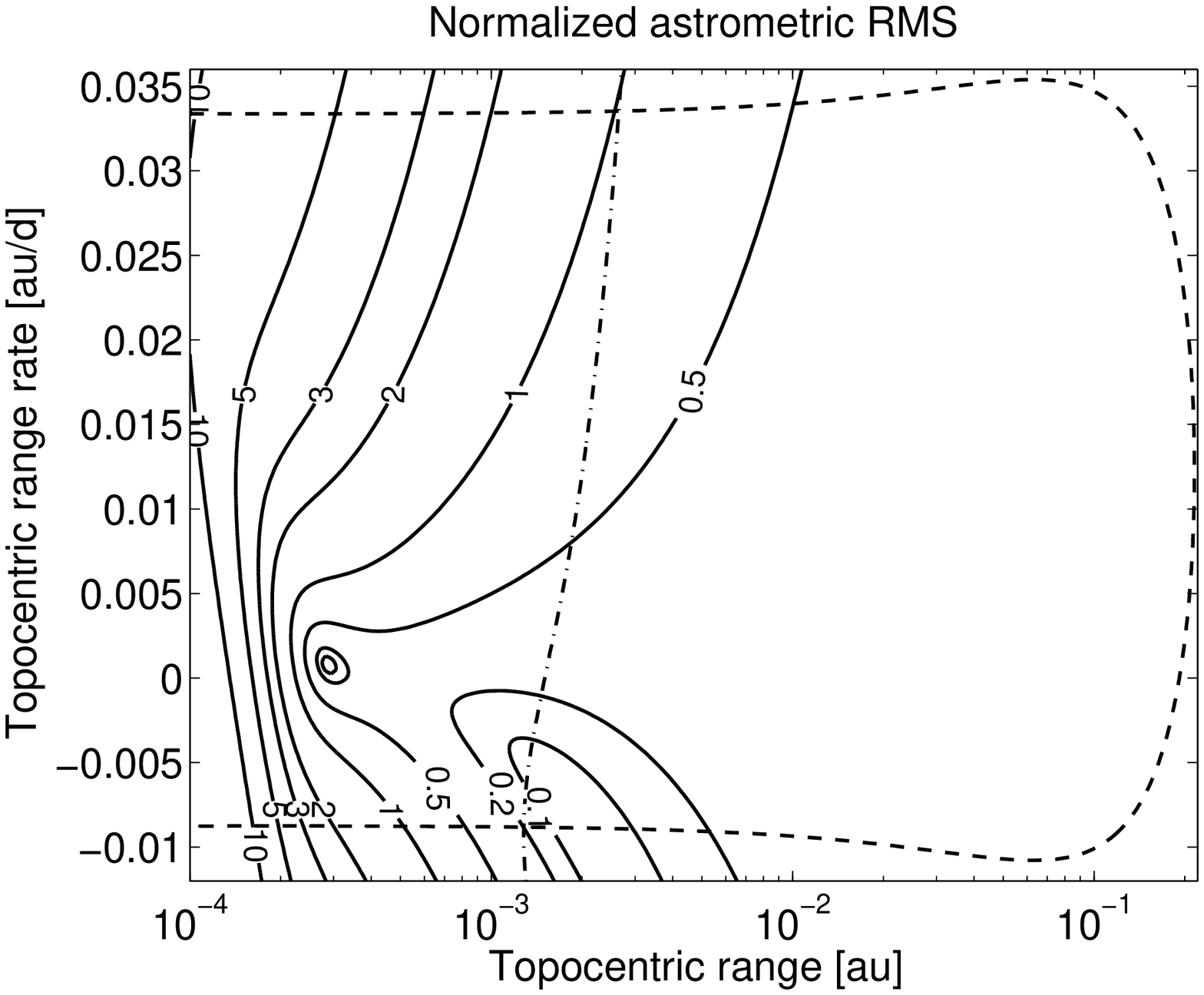}
  \includegraphics[width=0.5\textwidth]{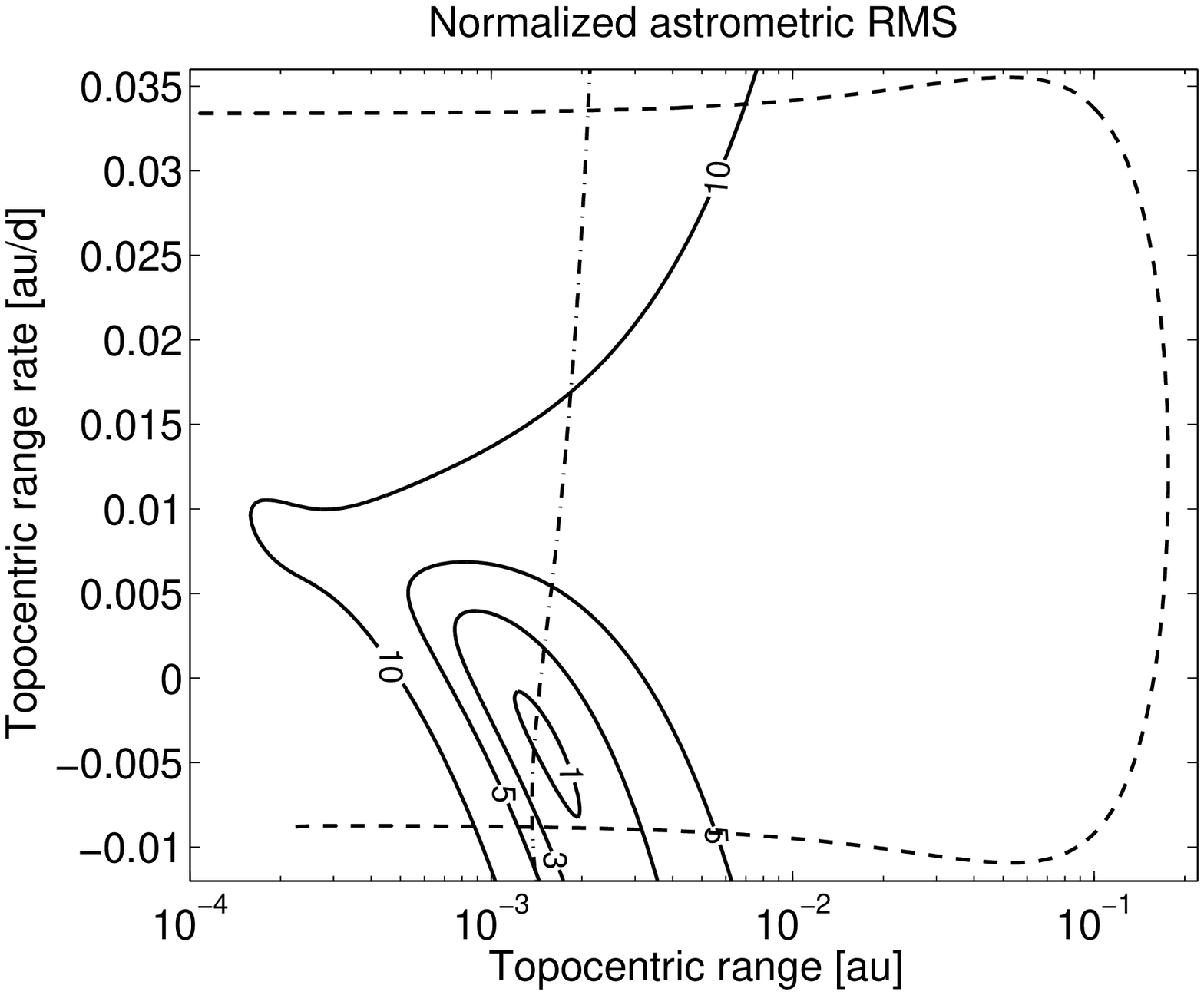}}
\caption{Normalized RMS as a function of $(\rho, \dot\rho)$ for
  Hayabusa. Left panel is for the first tracklet (three observations
  over 7 min), right panel is for the first two tracklets (total of
  seven observations over 34 min). Astrometric weights are at
  1.0$''$. The dashed curve corresponds to parabolic heliocentric
  orbits and the dash-dotted line to grazing impacts.}\label{f:hayabusa_rms}
\end{figure}

The reentry of Hayabusa took place on 2010 June
13. \citet{hayabusa_obs} report ground-based observations acquired
during the reentry phase, which ended with a landing in the South
Australia outback. We used these observations for applying systematic
ranging technique to Hayabusa. Figure~\ref{f:hayabusa_rms} shows the
normalized RMS as a function of $(\rho, \dot\rho)$ by using the first
tracklet of data (three observations) or the full available dataset
(six observations). Table~\ref{t:examples} gives the impact
probability as a function of the number of observations. With a
uniform prior the initial impact probability is $2.7\times 10^{-3}$
and increases to 17\% with the second tracklet. With the
population-based prior the initial impact probability is
$2.0\times 10^{-3}$ and increases to 2\% with the second tracklet. On
the other hand, with the use of Jefrreys' prior we immediately have a
94\% impact probability that decreases to 24\% with the second
tracklet. This counterintuitive behavior is due to the presence of two
local RMS minima with the first tracklet, one of which clearly
corresponds to an impact and is ruled out by the second tracklet.

Hayabusa 2 was launched on 2014 December
3.\footnote{http://global.jaxa.jp/press/2014/12/20141203\_h2af26.html}
After launch Hayabusa 2 was initially put on a low Earth parking
orbit, until a maneuver moved it onto a ballistic trajectory leaving
the Earth. On its way out, on 2014 December 6, Hayabusa 2 happend to
be in the field of view of the Pan-STARRS PS1 survey \citep{ps1},
which collected three observations of the spacecraft over 43 min. The
corresponding astrometry was submitted to the Minor Planet Center,
which, since the observations were not recognized as belonging to the
Hayabusa 2 spacecraft, put the object on the NEO Confirmation Page
\footnote{http://www.minorplanetcenter.net/iau/NEO/ToConfirm.html}
with the temporary designation P10gJfF. The following day, 2014
December 7, Pan-STARRS obtained three additional observations from
which it became clear that the object was Hayabusa
2. Table~\ref{t:examples} shows how systematic ranging recognizes
Hayabusa 2 as potential impactor from the first three observations
with an impact probability probability of few to ten percents, depending on the data
weights and the prior used. Then, the subsequent observations better
constrain the orbit and lower the impact probability since the
ballistic trajectory does not correspond to an impact but to the
post-injection escape orbit.

\subsection{2004~AS$_1$}
Asteroid 2004~AS$_1$ was discovered by LINEAR \citep{linear} on 2004
January 13 \citep{mpec_as1}. The four discovery observations span 71
min and show an
unusually high curvature thus suggesting that 2004~AS$_1$ was close to
the Earth and that an impact was possible. Follow-up observations
later showed that the object was much farther than originally thought,
thus ruling out any impact, and that the observed curvature was due to
large astrometric errors in the discovery observations.

\citet{virt_as1} employed the statistical ranging technique
\citep{stat_ran} to analyze the 2004~AS$_1$ case. They found an 18\%
impact probability with 1.0$''$ weights and concluded that
such a large impact probability was mostly due to the
poor quality of the discovery night observations.

\begin{figure}
\centerline{\includegraphics[width=0.5\textwidth]{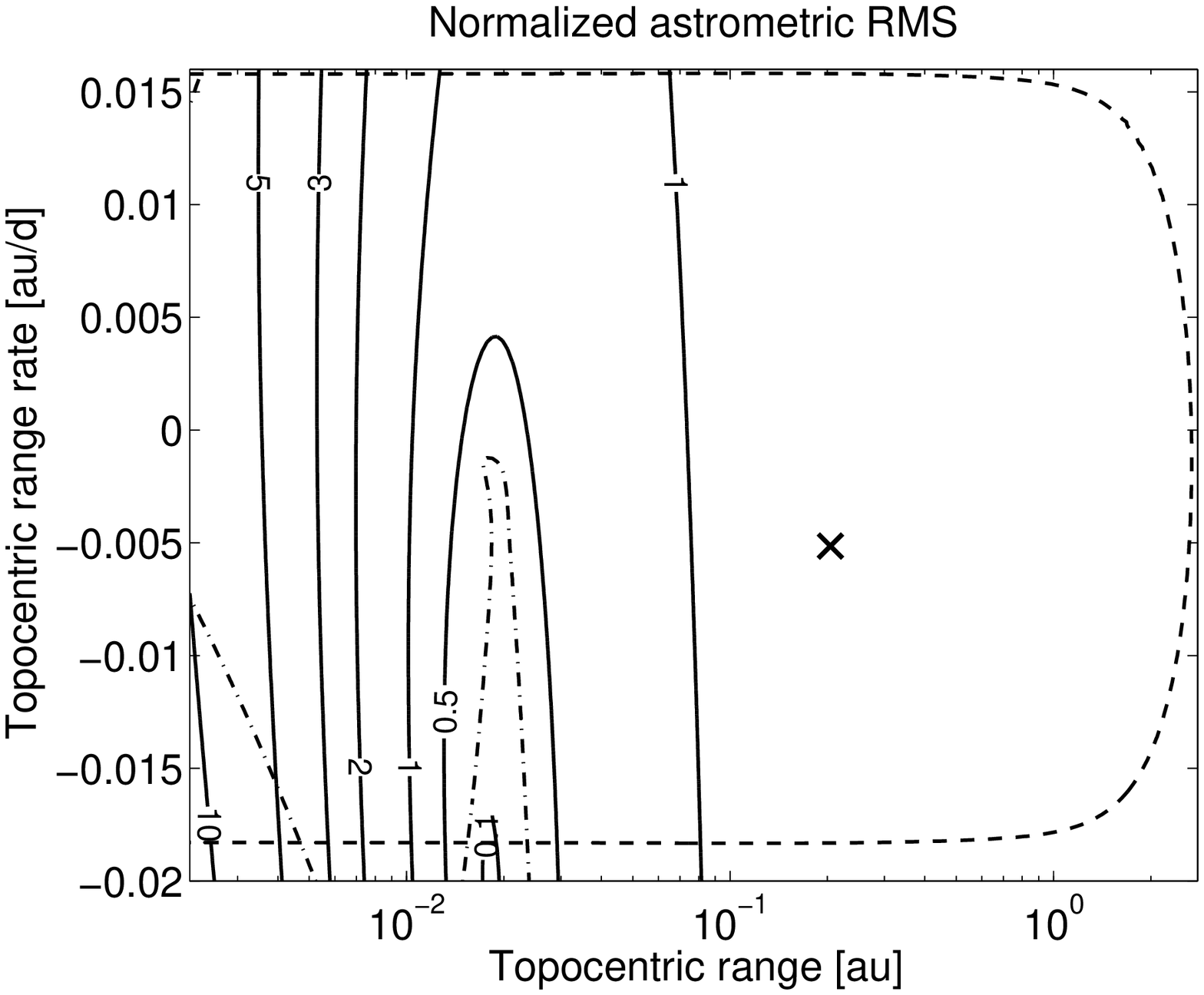}
\includegraphics[width=0.5\textwidth]{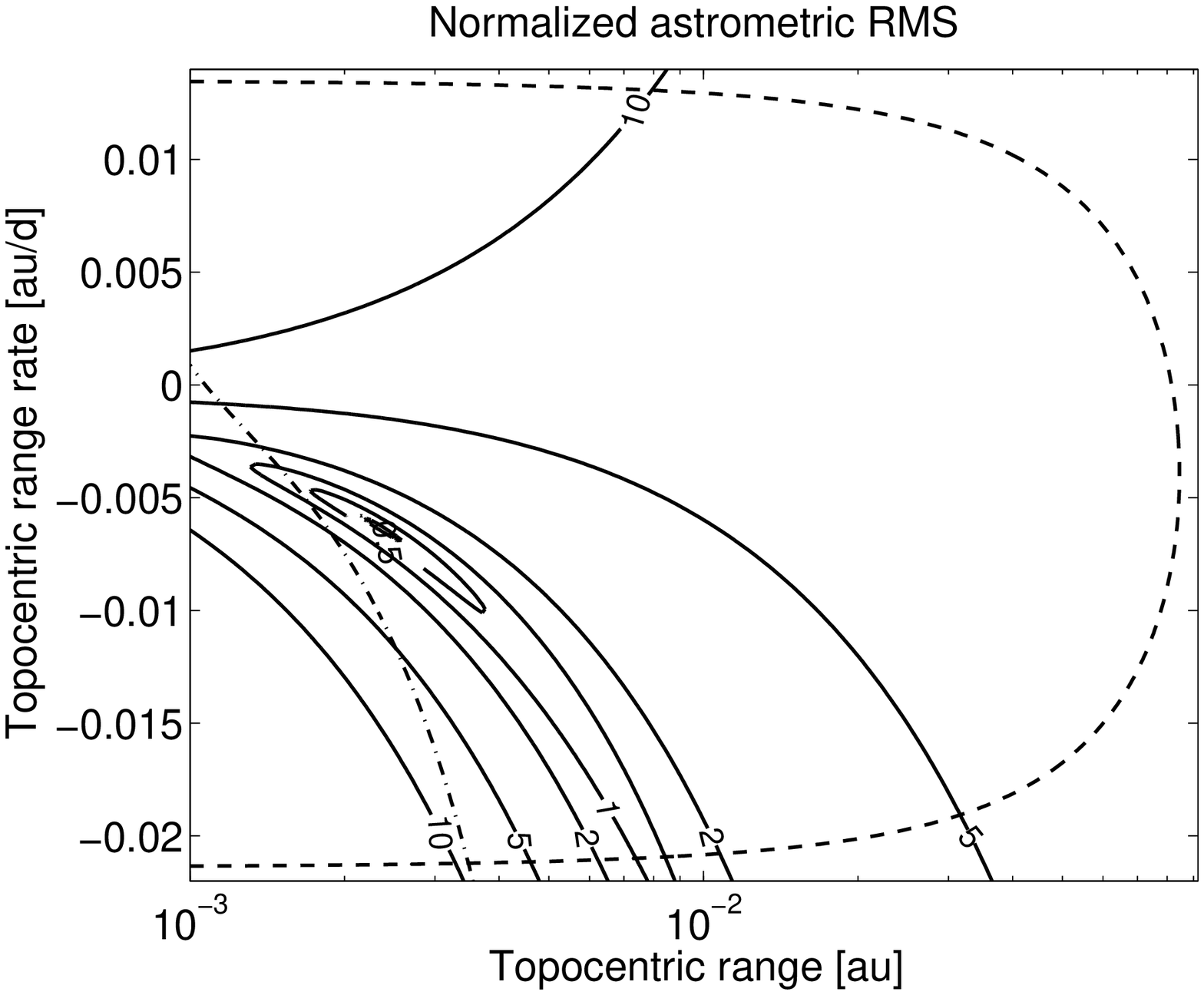}}
\caption{Normalized RMS as a function of $(\rho, \dot\rho)$ for
  2004~AS$_1$ (left panel) and 2004~FU$_{162}$ (right
  panel). Astrometric weights are at 1.0$''$.  The dashed curve
  corresponds to parabolic heliocentric orbits and the dash-dotted
  line to grazing impacts. For 2004~AS$_1$ the cross marks the true
  solution.}\label{f:04_rms}
\end{figure}

The left panel of Fig.~\ref{f:04_rms} shows the application of
systematic ranging to 2004~AS$_1$. The astrometric observations are
weighted at 1.0$''$. The RMS favors an impact solution while the true
solution (marked with a cross) correspond to larger astrometric
errors. The impact probabilities computed for the different choices of
the prior distribution are of the order of 10\%, see
Table~\ref{t:examples}. Decreasing the weights at 1.5$''$ gives
significantly lower probabilities for the uniform and population-based
priors, while with Jefrreys' prior we still have a 12\% impact
probability.

As shown by Table~\ref{t:pval}, we computed the $p$-value of the true
solution, i.e., the fraction of less likely orbital solutions, for the
different choices of the prior and different data weights. Jeffreys'
prior produces very small $p$-values thus indicating that the true
solution is deemed as extremely unlikely by the computed probability
distribution. For the uniform distribution we have 3\% $p$-value with
1$''$ weights, which is due to the large errors in the astrometric
observations. As a matter of fact, relaxing the weights increases the
$p$-value to 22\%. On the other hand, the population-based prior compensates
for the large astrometric errors since impacting solutions with
low RMS are less likely in the population model than non-impacting solutions
with a higher RMS. The $p$-values are 13.6\% and 38.7\% depending on
the data weights.

\begin{table}[t]
\begin{center}
\begin{tabular}{|lc|ccc|}
  \hline
  Object & Weights & \multicolumn{3}{c|}{$p$-value}\\
         & & Jeffreys & Uniform & Population\\
  \hline 
  2004~AS$_1$ & 1.0$''$ & $3.1\times 10^{-5}$ & $2.6\times 10^{-2}$
 & $1.4\times 10^{-1}$\\
  2004~AS$_1$ & 1.5$''$ & $4.0\times 10^{-4}$ & $2.2\times 10^{-1}$
 & $3.9\times 10^{-1}$\\
  \hline
  2015~CV & 0.5$''$ & $5.0\times 10^{-5}$ & $4.9\times 10^{-1}$
 & $4.2\times 10^{-1}$\\
  2015~CV & 1.0$''$ & $1.3\times 10^{-5}$ & $4.9\times 10^{-1}$
 & $4.2\times 10^{-1}$\\
  \hline
  2015~BU$_{92}$ & 0.5$''$ & $1.1\times 10^{-5}$ & $9.1\times 10^{-1}$
 & $8.8\times 10^{-1}$\\
  2015~BU$_{92}$ & 1.0$''$ & $2.3\times 10^{-6}$ & $9.1\times 10^{-1}$
 & $8.9\times 10^{-1}$\\
  \hline
\end{tabular}
\end{center}
\caption{$p$-values for some of the examples considered in
  Sec.~\ref{s:examples} for different choices of $f_{prior}$ and different data weights.}
\label{t:pval}
\end{table}

The conclusion fpr 2004~AS$_1$ is that the discovery tracklet,
weighted at 1$''$, yields impact probabilities $\sim10\%$ even as the
trueth is compatible with the data at the 3--14\% level. This outcome
is a manifestation of the large astrometric errors in the data.

\subsection{2004~FU$_{162}$}
Asteroid 2004~FU$_{162}$ was also discovered by LINEAR on 2004 March
31 \citep{mpec_fu162}.  \citet{sys_ran_steve} first analyzed this
object and estimated that there was a 0.3\% probability of an impact and
that the object was likely to pass within 10 Earth radii.

The right panel of Fig.~\ref{f:04_rms} shows our implementation of
systematic ranging, while Table~\ref{t:examples} reports the impact
probabilities for different weights and prior distributions. Again,
Jeffreys' prior results in very high impact probabilities while the
population-based prior gives the lowest impact chances. For 1$''$
weights the impact probability is 0.7\% -- 0.9\% with the uniform and
the population based priors. An Earth encounter within 10 Earth radii
is assured with all three priors, thus confirming the findings of
\citet{sys_ran_steve}.

\subsection{2015~CV \& 2015~BU$_{92}$}
\begin{figure}
\centerline{\includegraphics[width=0.5\textwidth]{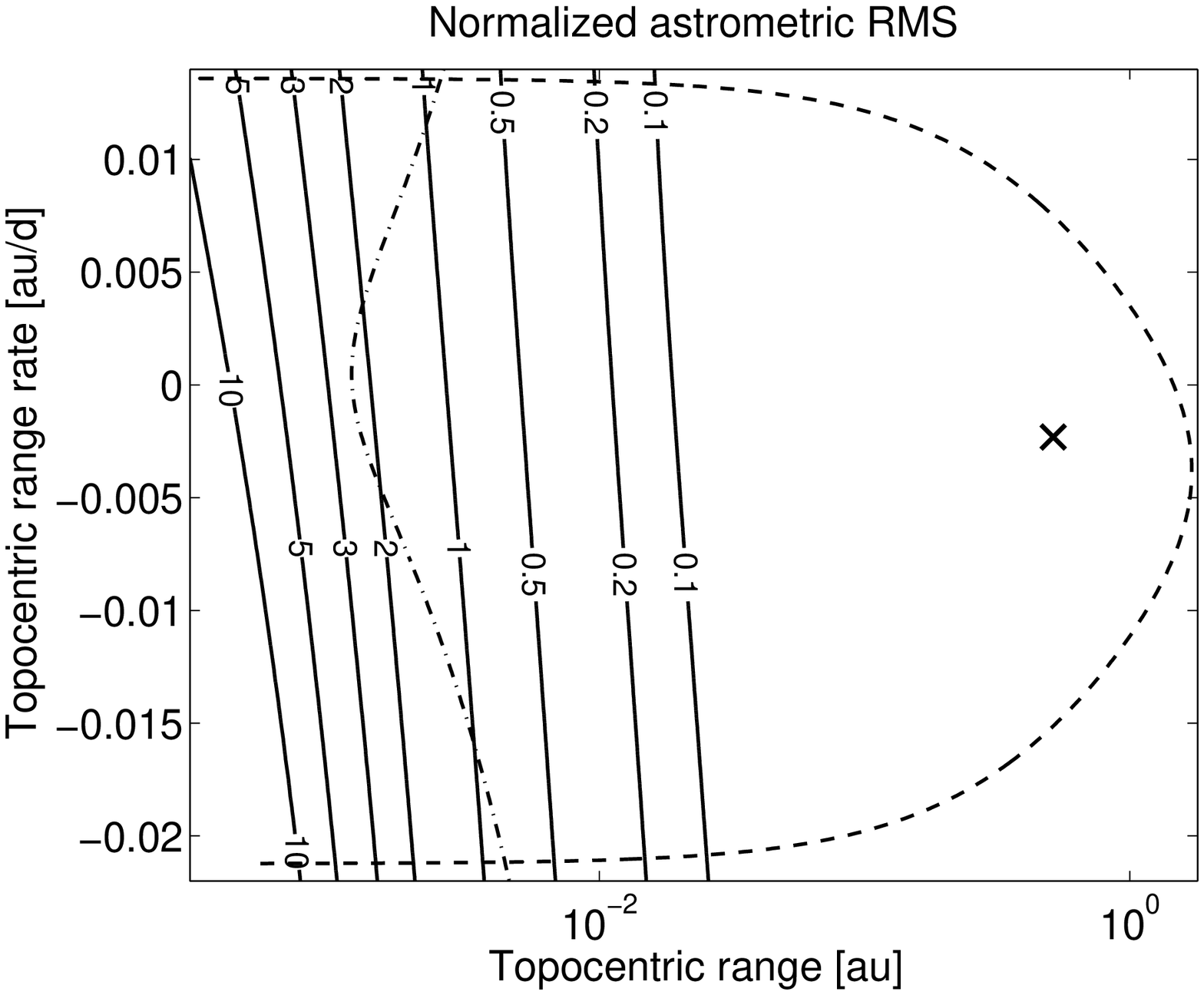}
\includegraphics[width=0.5\textwidth]{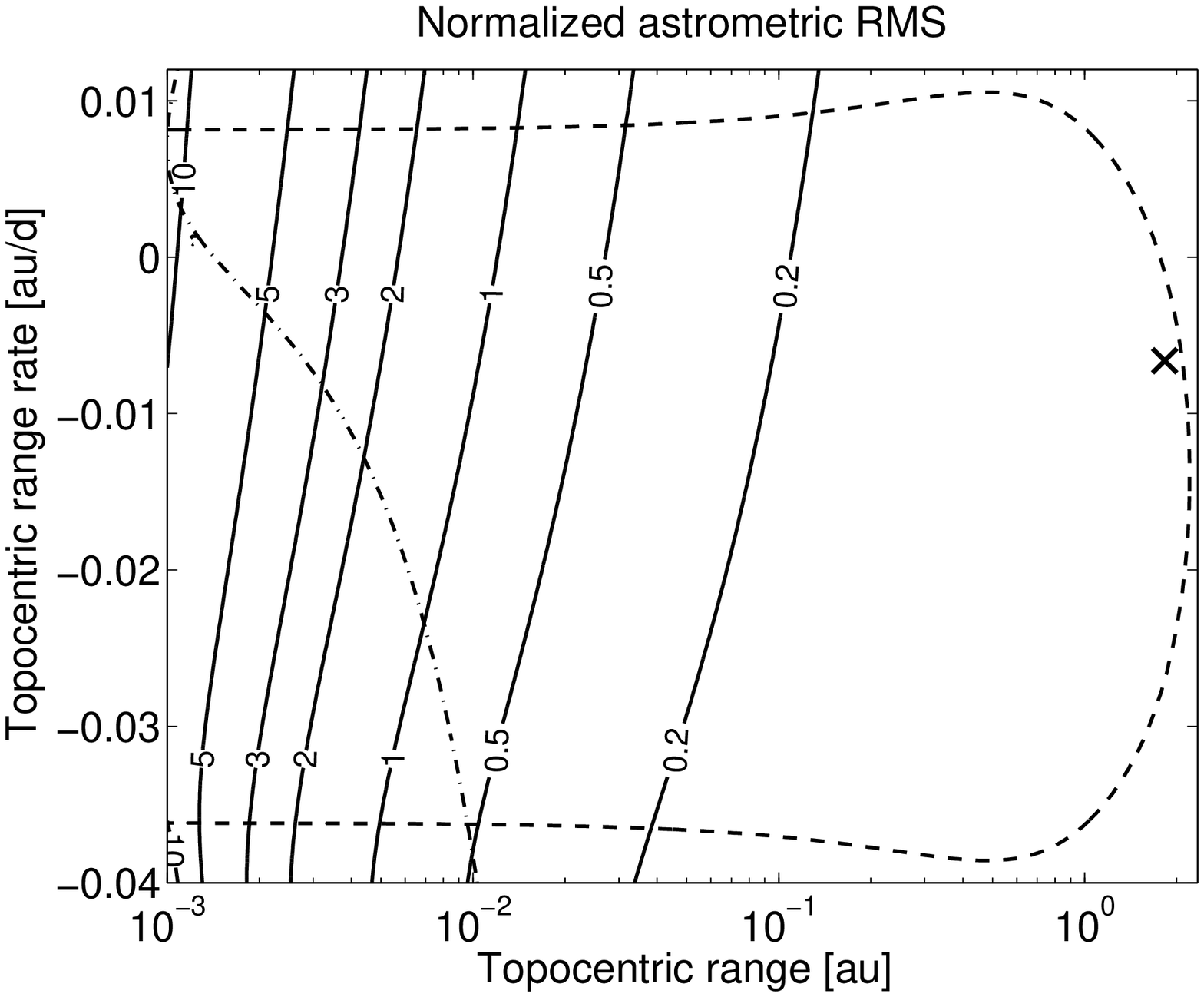}}
\caption{Normalized RMS as a function of $(\rho, \dot\rho)$ for
  2015~CV (left panel) and 2015~BU$_{92}$ (right panel). Astrometric
  weights are at 0.5$''$. The dashed curve corresponds to parabolic
  heliocentric orbits and the dash-dotted line to grazing
  impacts. Crosses mark the true solutions.}\label{f:non_imp_rms}
\end{figure}
The previous examples show how systematic ranging is capable of
quickly identifying actual and potential impactors. However, there is
the risk of having significant impact probabilities ($10^{-3}$ or
higher) for a large number of objects, which would make it difficult
to distinguish interesting objects from less remarkable ones. Here we
show the example of two main belt objects that were recently posted on
the NEO Confirmation Page when the initial observations were
compatible with an NEO orbit: 2015~CV and 2015~BU$_{92}$.

2015~CV was discovered by the Catalina Sky survey, Mt. Lemmon station,
on 2015 February 10, while 2015~BU$_{92}$ was discovered by Spacewatch
\citep{spacewatch}, Kitt Peak observatory, on 2015 January 16
\citep{mpec_bu92}. Figure~\ref{f:non_imp_rms} shows the application of
systematic ranging to these two asteroids.  For both objects the
region with reasonable astrometric errors extends from impacting
solutions to more than 1 au from the observer.

As shown by Table~\ref{t:examples} and Table~\ref{t:pval}, in both
cases Jeffreys' prior provides significant impact probabilities and
extremely low $p$-values for the true solutions. In fact, Jeffreys'
prior is proportional to the partials of the astrometric residuals and
the residuals do not change much for large topocentric distances while
they are very sensitive when the object is close to the
observer. Therefore, Jeffreys' prior tends to favor small topocentric
distances even if the astrometric errors are larger. On the other hand
the uniform and the population-based priors give much more reasonable
probabilities and $p$-values.


\section{Discussion and future work}
Small asteroids are expected to strike the Earth on a regular
basis. For instance, according to \citet{impact_frequency} impacts
with 10 m sized objects take place every few years. Since such small
asteroids are likely to be observed, if at all, only shortly before
impact, it is important to be capable of an early recognition of
hazardous objects.

The shortness of the observation arcs typically available for these
objects leads to critical degeneracies in the orbit estimation process
and limits the applicability of the standard impact monitoring systems
\citep{milani_im}. The systematic ranging technique presented in this
paper relies on Bayesian inversion theory, which overcomes these
degeneracies and results in a posterior distribution on the orbit phase space.

A critical point is the choice of a prior distribution for the poorly
constrained space of topocentric range and range rate. This choice is
far from trivial and indeed is somewhat arbitrary. We analyzed three
options: Jeffreys' prior, a uniform prior, and a prior distribution
based on the asteroid population model.

Jeffreys' prior does a good job at assigning high impact probabilities
to actual impactors. However, this prior produces high impact
probabilities also for less remarkable objects such as main belt
asteroids. The reason for this behavior is that Jeffreys' prior favors
regions in the phase space that are closer to the observer. As a
matter of fact, Jeffreys' prior is proportional to the partials of the
residuals with respect to topocentric range and range rate, which are
large for small topocentric distances and vanish at large
distances. Moreover, in the non-impacting examples we analyzed we
found that the true solution was often deemed as extremely unlikely by
the posterior distribution obtained starting from Jeffreys'
prior. Therefore, we avoid the use of Jeffreys' prior for systematic
ranging.

The prior based on the asteroid orbit and size distribution represents
the highest fidelity and most complex choice. However, it is worth
pointing out that a population model is necessarily limited by the
finite number of samples and is therefore affected by Poisson
statistics errors. In particular, small impactors might have peculiar
orbits that are deemed unlikely by the model, which could be too
much of a risk when the main goal is to identify potential impactors.

A uniform distribution represents a compromise between simplicity and
reliability. Though the corresponding posterior distribution and
impact probabilities are not as rigorous as the one resulting from the
population-based prior, we found that a uniform prior is effective at
flagging potential impactors and discriminating between hazardous
asteroids and more distant ones, which is the most important task to
be achieved by systematic ranging. Another advantage of the uniform
prior is that it has little prejudice on the actual orbit.

To complement the standard impact monitoring performed by JPL's
Sentry\footnote{http://neo.jpl.nasa.gov/risk} and the University of
Pisa's NEODyS,\footnote{http://newton.dm.unipi.it/neodys} we have
implemented an automated system to continually analyze the objects on
the Minor Planet Center's NEO Confirmation Page with systematic
ranging. This system quickly computes an impact probability to
identify potential impactors and notifies us of interesting
results. In such cases, we have communicated with discoverers and
other observers to solicitate a quality control of the reported astrometry
and for acquiring follow-up observations, which are necessary to
decrease the orbital uncertainties and conclusively determine if the
impact is going to take place.  The examples of 2008~TC$_3$ and
2014~AA, i.e., the only two asteroids discovered before striking the
Earth, suggest that objects with an impact probability above $10^{-3}$
should be prioritized for prompt follow-up. In recent experience this
happens a few to several times a month.

Among the possible additional applications of the systematic ranging
technique presented in this paper, one could be the identification
potential radar targets or mission-accessible targets, e.g., for the
NASA's Asteroid Redirect Mission \citep{arm}.

\section*{Acknowledgments}
We thank T.~B. Spahr for his advice on the astrometric weights of
Table~\ref{t:weights}.

Part of this research was conducted at the Jet Propulsion Laboratory,
California Institute of Technology, under a contract with NASA.

Copyright 2015 California Institute of Technology.

\end{document}